\documentclass[11pt]{article} 
\usepackage{epsfig} 
\usepackage{cite}  
\def\e{\epsilon}

\newcommand{\be}{\begin{equation}}
\newcommand{\eps}{\epsilon}
\newcommand{\dps}{\displaystyle}
\newcommand{\ee}{\end{equation}}
\newcommand{\bea}{\begin{eqnarray}}
\newcommand{\eea}{\end{eqnarray}}

\newcommand{\loopint}[1]{\int \!\!\! \frac{d^D #1}{\left(2\pi\right)^D}\!}
\newcommand{\ESGamma}{S_{\Gamma}}

\newcommand{\lk}{\left(}
\newcommand{\rk}{\right)}
\newcommand{\lek}{\left[}
\newcommand{\rek}{\right]}
\newcommand{\lp}{\left.}
\newcommand{\rp}{\right.}
\newcommand{\nnb}{\nonumber}
\newcommand{\ts}[2]{\textstyle{\frac{#1}{#2}}}

\newcommand{\pFq}[5]{\, \! _{#1} F_{#2}( #3 \, ; \, #4 \, ; \, #5 )}

\newcommand{\hs}[1]{\hspace*{#1 pt}}
\newcommand{\vs}[1]{\vspace*{#1 pt}}

\begin{document} 
\begin{titlepage} 
\vspace*{-1cm} 
\begin{flushright} 
ZU--TH 16/06\\
July 2006
\end{flushright} 
\vskip 3.5cm 

\begin{center} 
{\Large\bf Master Integrals for Massless Three-Loop Form Factors:\\[2mm]
One-Loop and Two-Loop Insertions}
\vskip 1.cm 
{\large  T.~Gehrmann}, {\large G.~Heinrich}, {\large T.~Huber} 
and {\large C.\ Studerus}
\vskip .7cm 
{\it Institut f\"ur Theoretische Physik, Universit\"at Z\"urich,
Winterthurerstrasse 190,\\ CH-8057 Z\"urich, Switzerland} 
\end{center} 
\vskip 2cm 

\begin{abstract} 
The three-loop form factors in massless QCD can be expressed as a 
linear combination of master integrals. Besides a number of master 
integrals which factorise into products of one-loop and two-loop integrals,
one finds 16 genuine three-loop integrals. Of these, six have the form of a
bubble insertion inside a one-loop or two-loop vertex integral. We compute 
all master integrals with these insertion topologies. 
\end{abstract} 
\vfill 
\end{titlepage} 
\newpage 

\section{Introduction}
The vertex functions of a virtual photon coupling to a quark-antiquark 
pair (quark form factor) and of a Higgs boson coupling to two 
gluons through an effective coupling (gluon form factor) are 
the simplest diagrams containing infrared divergencies in 
higher orders in massless quantum field theory. These form 
factors appear in a wide variety of applications: they can be used to 
predict the infrared pole structure of multi-leg amplitudes at a given 
order~\cite{catani,sterman} and to extract resummation 
coefficients~\cite{moch1}, 
and they make up the purely virtual corrections to a number of collider 
reactions (Drell-Yan process, Higgs production and decay, deep inelastic 
scattering). 

In the past, two-loop corrections to the massless quark~\cite{vanneerven} and 
gluon~\cite{harlander,ravindran} form
factors were computed in dimensional 
regularisation in $D=4-2\e$ dimensions to order $\e^0$. 
 Two-loop corrections to this 
order were also obtained for massive quarks~\cite{breuther}. The 
massless two-loop 
form factors were extended to all orders in $\e$ in~\cite{ghm}, and 
three-loop form factors to order $\e^{-1}$ (and $\e^0$ for 
fermion loop contributions) were computed in~\cite{moch1,moch2}. These 
three-loop results had an immediate application 
in the calculation of the N$^3$LO threshold-enhanced 
soft emission corrections~\cite{moch3}  
to the inclusive Drell-Yan and Higgs production cross section, 
demonstrating the perturbative stability at this order. 

In~\cite{moch1,moch2}, the form factors were inferred from the 
behaviour of the three-loop deep inelastic coefficient 
functions~\cite{moch4}; this 
procedure can not be easily extended to yield also all the finite 
terms. Instead, one can turn to the more conventional approach of  
computing  multi-loop Feynman amplitudes, which   proceeds through a 
reduction~\cite{chet,laporta,gr} of all  Feynman 
integrals appearing in the form factors to a small set of master integrals. 
The reduction is purely algebraic and can be automated using 
computer algebra methods~\cite{laporta,air}. The master integrals 
take the form of a Laurent series in $\e$, and they must be 
computed to a given order in $\e$, typically specified by the
transcendentality of the coefficients. The finite part of the
three-loop form factors requires transcendentality six, i.e.\ 
coefficients containing terms up to $\pi^6$ or $\zeta_3^2$.  

In this letter, we identify all master integrals needed for the 
three-loop form factors
in Section~\ref{sec:mi}. Many of these are products of integrals with 
one or two loops, or three-loop propagator integrals. Among the remaining 
genuine three-loop vertex integrals, several contain one-loop or two-loop 
propagator insertions. We describe how the Laurent expansion of these 
insertion topologies can be obtained either analytically or numerically 
in Section~\ref{sec:comp}, and list the results for them in 
Section~\ref{sec:results}. 
Finally, Section~\ref{sec:conc} contains our conclusions and an outlook.

\section{Master integrals for three-loop form factors}
\label{sec:mi}

The topologies of the  master integrals relevant 
to three-loop form factors 
can be inferred from two-particle cuts of the
master integrals of massless four-loop off-shell propagator 
integrals (massless four-loop two-point functions). The master 
integrals of these massless four-loop two-point functions were 
identified in~\cite{baikov} and subsequently used in the calculation 
of the scalar $R$-ratio~\cite{bck}. Analytical expressions for these 
integrals are, however, not available in the literature. Since each 
two-particle cut is in general only one of several (two-, 
three-, four- and five-particle) cuts, knowledge of these two-point 
master integrals would not facilitate the calculation of the 
master integrals for the three-loop form factors.

\begin{figure}[p]
    \vs{5}
    \hs{31}
  \begin{tabular}{cccc}
  \parbox{3.3cm}{
\begin{picture}(0,0)
\thicklines
\put(-30,0){\vector(1,0){17}}
\put(-13,0){\line(1,0){13}}
\put(0,0){\line(3,2){40}}
\put(0,0){\line(3,-2){40}}
\put(0,0){\vector(3,2){37}}
\put(0,0){\vector(3,-2){37}}
\put(24,-16){\line(0,1){32}}
\qbezier[80](24,-15.9)(34,0)(24,16)
\qbezier[80](24,-16.1)(12.5,0)(24,16)
\end{picture}\\ \vs{30} \\ $A_{5,1}$}
  & 
\parbox{3.3cm}{
\begin{picture}(0,0)
\thicklines
\put(-30,0){\vector(1,0){17}}
\put(-13,0){\line(1,0){13}}
\put(0,0){\line(3,2){40}}
\put(0,0){\line(3,-2){40}}
\put(0,0){\vector(3,2){37}}
\put(0,0){\vector(3,-2){37}}
\put(24,-16){\line(0,1){32}}
\qbezier[80](24,-16)(34,0)(24,16)
\qbezier[80](0,0)(8,-20)(24,-16)
\end{picture}\\ \vs{30} \\ $A_{5,2}$}
  & 
\parbox{3.3cm}{
\begin{picture}(0,0)
\thicklines
\put(-30,0){\vector(1,0){17}}
\put(-13,0){\line(1,0){13}}
\put(0,0){\line(3,2){40}}
\put(0,0){\line(3,-2){40}}
\put(0,0){\vector(3,2){37}}
\put(0,0){\vector(3,-2){37}}
\qbezier[60](24,16.3)(18,12)(24,0)
\qbezier[60](24,-16.3)(18,-12)(24,0)
\qbezier[60](24,16.3)(30,12)(24,0)
\qbezier[60](24,-16.3)(30,-12)(24,0)
\end{picture}\\ \vs{30} \\ $A_{6,1}$}
  &
\parbox{3.3cm}{
\begin{picture}(0,0)
\thicklines
\put(-30,0){\vector(1,0){17}}
\put(-13,0){\line(1,0){13}}
\put(0,0){\line(3,2){40}}
\put(0,0){\line(3,-2){40}}
\put(0,0){\vector(3,2){37}}
\put(0,0){\vector(3,-2){37}}
\put(27,-18){\line(0,1){36}}
\put(26.8,-17.8){\line(-1,3){5.9}}
\put(26.8,17.8){\line(-1,-3){5.9}}
\put(0,0){\line(1,0){20.8}}
\end{picture}\\ \vs{30} \\ $A_{6,2}$} \\ \\ \\ \\ \\
  \end{tabular}
  \hs{50}
  \begin{tabular}{ccc}
  \parbox{4.5cm}{
\begin{picture}(0,0)
\thicklines
\put(-50,0){\vector(1,0){27}}
\put(-23,0){\line(1,0){23}}
\put(0,0){\line(3,2){60}}
\put(0,0){\line(3,-2){60}}
\put(0,0){\vector(3,2){57}}
\put(0,0){\vector(3,-2){57}}
\put(46.2,-30.8){\line(0,1){61.6}}
\put(18,12){\line(2,-3){28.7}}
\qbezier[80](17.6,12.9)(20,40)(46,31)
\end{picture}\\ \vs{40} \\ $A_{6,3}$}
 &
\parbox{4.5cm}{
\begin{picture}(0,0)
\thicklines
\put(-50,0){\vector(1,0){27}}
\put(-23,0){\line(1,0){23}}
\put(0,0){\line(3,2){60}}
\put(0,0){\line(3,-2){60}}
\put(0,0){\vector(3,2){55}}
\put(0,0){\vector(3,-2){55}}
\put(24,-16){\line(2,5){17.4}}
\put(42,-28){\line(-1,2){10.5}}
\put(27,2){\line(-1,2){6}}
\qbezier[60](0,0)(0,26)(20.7,14)
\end{picture}\\ \vs{40} \\ $A_{7,1}$}
  &   
  \parbox{4.5cm}{
\begin{picture}(0,0)
\thicklines
\put(-50,0){\vector(1,0){27}}
\put(-23,0){\line(1,0){23}}
\put(0,0){\line(3,2){60}}
\put(0,0){\line(3,-2){60}}
\put(0,0){\vector(3,2){55}}
\put(0,0){\vector(3,-2){55}}
\put(24,-16){\line(2,5){17.4}}
\put(42,-28){\line(-1,2){10.5}}
\put(27,2){\line(-1,2){6}}
\qbezier[60](20.7,14)(20,40)(41,27.7)
\end{picture}\\ \vs{40} \\ $A_{7,2}$}    \\ \\ \\ \\ \\
\parbox{4.5cm}{
\begin{picture}(0,0)
\thicklines
\put(-50,0){\vector(1,0){27}}
\put(-23,0){\line(1,0){23}}
\put(0,0){\line(3,2){60}}
\put(0,0){\line(3,-2){60}}
\put(0,0){\vector(3,2){57}}
\put(0,0){\vector(3,-2){57}}
\put(46.2,-30.8){\line(0,1){61.6}}
\put(18,12){\line(2,-3){28.7}}
\put(18,-12){\line(0,1){24}}
\end{picture}\\ \vs{40} \\ $A_{7,3}$}
 &
 \parbox{4.5cm}{
\begin{picture}(0,0)
\thicklines
\put(-50,0){\vector(1,0){27}}
\put(-23,0){\line(1,0){23}}
\put(0,0){\line(3,2){60}}
\put(0,0){\line(3,-2){60}}
\put(0,0){\vector(3,2){57}}
\put(0,0){\vector(3,-2){57}}
\put(18,-12){\line(0,1){24}}
\put(18,-12){\line(2,3){6.7}}
\put(28,3){\line(2,3){18.5}}
\put(18,12){\line(2,-3){28.7}}
\end{picture}\\ \vs{40} \\ $A_{7,4}$}
&
\parbox{4.5cm}{
\begin{picture}(0,0)
\thicklines
\put(-50,0){\vector(1,0){27}}
\put(-23,0){\line(1,0){23}}
\put(0,0){\line(3,2){60}}
\put(0,0){\line(3,-2){60}}
\put(0,0){\vector(3,2){57}}
\put(0,0){\vector(3,-2){57}}
\put(46.2,-30.8){\line(0,1){61.6}}
\put(18,-12){\line(2,3){6.7}}
\put(28,3){\line(2,3){18.5}}
\put(18,12){\line(2,-3){28.7}}
\end{picture}\\ \vs{40} \\ $A_{7,5}$} \\ \\ \\ \\ \\
 \parbox{4.5cm}{
\begin{picture}(0,0)
\thicklines
\put(-50,0){\vector(1,0){27}}
\put(-23,0){\line(1,0){23}}
\put(0,0){\line(3,2){64}}
\put(0,0){\line(3,-2){64}}
\put(0,0){\vector(3,2){60}}
\put(0,0){\vector(3,-2){60}}
\put(30,-20){\line(0,1){40}}
\put(30,-20){\line(-4,5){20.7}}
\put(18,-5){\line(5,6){9}}
\put(34,14.2){\line(5,6){15.5}}
\end{picture}\\ \vs{40} \\ $A_{8,1}$}
  &   
  \parbox{4.5cm}{
\begin{picture}(0,0)
\thicklines
\put(-50,0){\vector(1,0){27}}
\put(-23,0){\line(1,0){23}}
\put(0,0){\line(3,2){60}}
\put(0,0){\line(3,-2){60}}
\put(0,0){\vector(3,2){55}}
\put(0,0){\vector(3,-2){55}}
\put(42,-28){\line(0,1){56}}
\put(42,-8){\line(-2,1){30.9}}
\put(42,-28){\line(-1,2){11}}
\put(27,2){\line(-1,2){5.9}}
\end{picture}\\ \vs{40} \\ $A_{8,2}$}
&
\parbox{4.5cm}{
\begin{picture}(0,0)
\thicklines
\put(-50,0){\vector(1,0){27}}
\put(-23,0){\line(1,0){23}}
\put(0,0){\line(3,2){60}}
\put(0,0){\line(3,-2){60}}
\put(0,0){\vector(3,2){55}}
\put(0,0){\vector(3,-2){55}}
\put(24,-16){\line(0,1){32}}
\put(42,-28){\line(0,1){56}}
\put(24,0){\line(1,0){18}}
\end{picture}\\ \vs{40} \\ $A_{9,1}$} \\ \\ \\ \\ \\
\parbox{4.5cm}{
\begin{picture}(0,0)
\thicklines
\put(-50,0){\vector(1,0){27}}
\put(-23,0){\line(1,0){23}}
\put(0,0){\line(3,2){64}}
\put(0,0){\line(3,-2){64}}
\put(0,0){\vector(3,2){60}}
\put(0,0){\vector(3,-2){60}}
\put(18,-12){\line(0,1){24}}
\put(30,-20){\line(0,1){40}}
\put(18,-5){\line(5,6){9}}
\put(34,14.2){\line(5,6){15.5}}
\end{picture}\\ \vs{40} \\ $A_{9,2}$}
&
 \parbox{4.5cm}{
\begin{picture}(0,0)
\thicklines
\put(-50,0){\vector(1,0){27}}
\put(-23,0){\line(1,0){23}}
\put(0,0){\line(3,2){60}}
\put(0,0){\line(3,-2){60}}
\put(0,0){\vector(3,2){55}}
\put(0,0){\vector(3,-2){55}}
\put(24,-16){\line(0,1){13}}
\put(24,4){\line(0,1){12}}
\put(42,-28){\line(0,1){56}}
\put(42,-8){\line(-2,1){30.9}}
\end{picture}\\ \vs{40} \\ $A_{9,3}$}
& 
 \parbox{4.5cm}{
\begin{picture}(0,0)
\thicklines
\put(-50,0){\vector(1,0){27}}
\put(-23,0){\line(1,0){23}}
\put(0,0){\line(3,2){60}}
\put(0,0){\line(3,-2){60}}
\put(0,0){\vector(3,2){56}}
\put(0,0){\vector(3,-2){56}}
\put(24,-16){\line(0,1){13.4}}
\put(42,-28){\line(0,1){25.6}}
\put(23.8,-2.6){\line(1,0){18.4}}
\put(42,-2.5){\line(-2,3){14}}
\put(24.1,-2.6){\line(2,3){7}}
\put(35.1,13.9){\line(2,3){11.3}}
\end{picture}\\ \vs{40} \\ $A_{9,4}$} \\ \\ \\
  \end{tabular}
    \vs{-20} 
  \caption{Three-loop master integrals with massless propagators. The incoming momentum is $q=p_1+p_2$. Outgoing lines are considered
  on-shell and massless, i.e. $p_1^2=p_2^2=0$.  \label{fig:3loopdiagrams}}
  \end{figure}
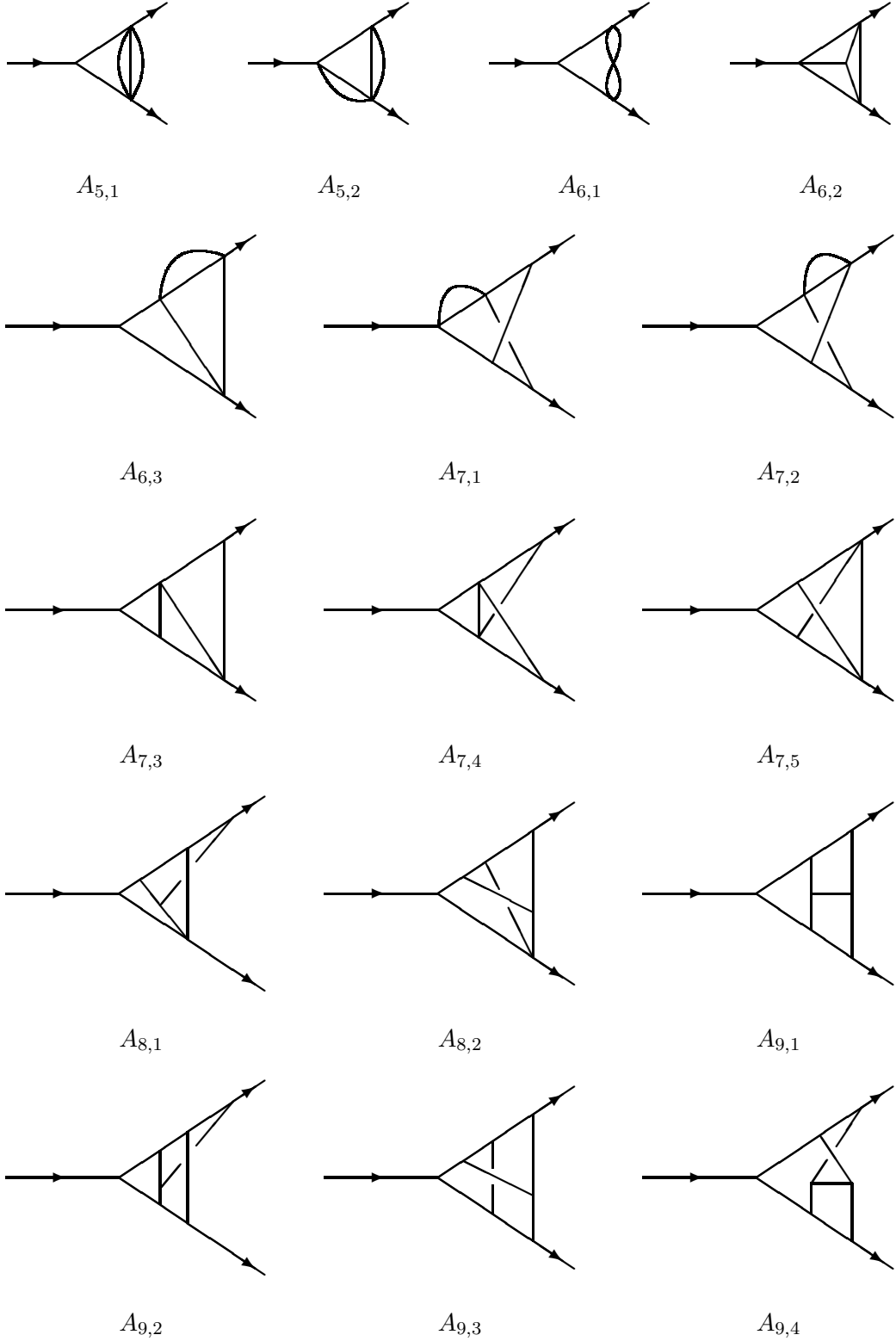

These master integrals can be classified into three types: (i) products 
of one-loop and two-loop vertex functions
with one off-shell and two on-shell legs, (ii) three-loop  two-point 
functions, (iii) three-loop vertex functions with one off-shell and 
two on-shell legs. Since the one-loop and two-loop vertex functions are 
known to all orders in $\e$~\cite{ghm}, all master integrals 
of type (i) can be obtained directly by expansion~\cite{hypexp,xsummer} of 
the all-orders 
results. Likewise, three-loop two-point functions
appearing in type (ii) 
are known to sufficiently high orders in $\e$~\cite{chet,bekavac} and are  
tabulated for example in the MINCER package~\cite{mincer}. The only
non-trivial master integrals for three-loop form factors are therefore of 
type (iii). The full set of these integrals is displayed in 
Figure~\ref{fig:3loopdiagrams}. Each topology contains only one master 
integral, which is chosen to be the scalar integral, with 
no loop momenta in the numerator and  with all propagators 
raised to unit power. Nevertheless, we will give the results for the
two-loop insertions for arbitrary propagator powers, 
see Section~\ref{sec:results}.
The topologies  $A_{5,2}$ and $A_{6,2}$ with some of the 
lines being massive have been calculated 
in~\cite{Marquard:2006qi}, where they enter the calculation of the 
three-loop matching coefficient of the heavy quark current.  

Among the master integrals of Figure~\ref{fig:3loopdiagrams}, the integrals 
$A_{5,1}$, $A_{5,2}$, $A_{6,1}$, $A_{6,3}$, $A_{7,1}$, $A_{7,2}$
are of special character, since they contain 
either a one-loop two-point insertion into a two-loop vertex integral 
($A_{6,3}$, $A_{7,1}$, $A_{7,2}$) or a two-loop (or one-loop times one-loop) 
two-point insertion into a one-loop vertex 
integral ($A_{5,1}$, $A_{5,2}$, $A_{6,1}$). These so-called insertion topologies 
are in general
simpler than the remaining genuine three-loop vertex integrals, since 
they can be obtained by computing a one-loop or two-loop vertex 
function with one or two propagators raised to a symbolic power. 
In the following, we describe the calculation of these insertion topologies. 

\section{Computational methods}
\label{sec:comp}
Three-loop vertex integrals with one off-shell and two on-shell legs
and massless propagators depend only on one kinematical scale: the mass $q^2$
of the off-shell leg. The dependence on this scale is given by the 
mass dimension of the integral, such that the coefficients of the  
Laurent expansion are constants, i.e. real numbers (which are in general of 
increasing transcendentality). Several techniques exist to compute such 
single-scale integrals. 

For all one-loop and two-loop insertion topologies considered here, we 
performed two independent calculations, using two different techniques:
evaluation in terms of hypergeometric series from Feynman parametrisation and
evaluation using sector decomposition. 

The Feynman parametrisation for the one-loop and two-loop vertex 
functions with symbolic powers on individual propagators results in a
multiple integral in the Feynman parameters. Depending on the topology, 
one has to integrate over at least two (one-loop vertex function) and 
at most five 
(non-planar two-loop vertex function) Feynman parameters. After
appropriately decomposing the integration region to avoid parametric 
singularities~\cite{kl}, and introducing supplementary regulators 
at intermediate stages, one can express the results of this 
integration in terms of hypergeometric functions of unit argument,
containing $\e$-dependent coefficients. These can be expanded in $\e$ 
using the {\tt Mathematica}~\cite{Mathematica} 
package {\tt HypExp}~\cite{hypexp} to yield the Laurent 
series of the master integrals. 

For many practical applications, and to verify the analytical results, 
it is sufficient to know the numerical 
values of the coefficients in the 
Laurent expansion of the master integrals to some finite order. 
These can be obtained using the sector decomposition technique.
 
The sector decomposition technique for the computation of multi-loop 
integrals is described in detail in~\cite{gudrun1,gudrun2}.
Using this technique, the Laurent expansions of all 
master integrals relevant to the three-loop form factors can be computed 
to any desired order, limited only by computation time. First applications 
to three-loop vertex integrals  
were presented already in~\cite{gudrun1}. 
The treatment of propagator powers different from unity 
is described in~\cite{gudrun2}.
The application of sector decomposition to the topologies 
$A_{6,3}$, $A_{7,1}$ and $A_{7,2}$ has been done in two different ways: \\
(a) by direct calculation of the three-loop topologies,  
(b) by calculating the two-loop diagram 
with $\eps$-dependent propagator powers resulting from 
integrating out the one-loop two-point insertion (corresponding to 
$I_5(\eps),\, I_6(\eps),\, J_6(\eps)$ in Section \ref{sec:results}). 
The analytical results for $A_{5,1}$, $A_{5,2}$ and $A_{6,1}$, 
given for general symbolic propagator powers $\nu_i$ in 
Section \ref{sec:results}, also have been verified
for some $\eps$-dependent $\nu_i$ values  by sector decomposition.

The computing time for a seven propagator graph like 
$A_{7,1}$ or $A_{7,2}$ up to order $\eps^0$ for a 
numerical precision better than 0.1\% is of the order 
of 20\,minutes on a 2.8 GHz PC, while 
the order $\eps$ term takes about 6\,hours. 
For a precision of 1\% the evaluation 
is about 10 times faster.

\section{Results for the insertion topologies}
\label{sec:results}

In this section we list the results we obtained for the three loop master integrals with insertion topology. The labelling of the diagrams
is according to Figure~\ref{fig:3loopdiagrams}. The results for the diagrams $A_{5,1}$, $A_{5,2}$, and $A_{6,1}$ can be given for arbitrary
propagator powers $\nu_i$. The values of the $\nu_i$ are assumed to be such that the arguments of all occurring $\Gamma$-functions are
different from
\begin{displaymath}
0, \, -1, \, -2 ,\ldots \; \; .
\end{displaymath}

In our first diagram, namely $A_{5,1}$, we label the powers of the 
sloped propagators (i.e. the ones attached to the off-shell leg) 
by $\nu_1$ and $\nu_2$, whereas $\nu_3$, $\nu_4$,
and $\nu_5$ are associated with the three propagators that form the twofold bubble insertion. The form of the
diagram immediately suggests that the result must be completely symmetric in $\{\nu_1,\nu_2\}$ as well as in $\{\nu_3,\nu_4,\nu_5\}$.
The calculation leads to
\bea
\dps A_{5,1}[\nu_i] &=& \loopint{k}\loopint{l}\loopint{r} \; \; \frac{1}{\lek(k+p_1)^2\rek^{\nu_1} \, \lek (k-p_2)^2\rek^{\nu_2}  
\, \lek l^2\rek^{\nu_3} \, \lek (k+l+r)^2 \rek^{\nu_4} \, \lek r^2\rek^{\nu_5}} \nnb \\
&&\nnb \\
&=&\frac{i \, (-1)^{1-N}}{(4\pi)^{3D/2}}  \, \lek -\lp q\rp^2-i \, \eta \rek^{3D/2-N} \; \frac{\Gamma(\ts{D}{2}-\nu_3) \,
\Gamma(\ts{D}{2}-\nu_4) \, \Gamma(\ts{D}{2}-\nu_5)}{\Gamma(\nu_1) \, \Gamma(\nu_2) \, \Gamma(\nu_3) \, \Gamma(\nu_4)\, \Gamma(\nu_5)}
\nnb\\
&&\nnb \\
&& \times \, \frac{\Gamma(N-\ts{3 \, D}{2}) \, \Gamma(\nu_{345}-D) \, \Gamma(\ts{3\,D}{2}- N + \nu_{1}) \,
\Gamma(\ts{3\,D}{2}-N+\nu_{2})}{\Gamma(\ts{3\,D}{2}-\nu_{345}) \, \Gamma(2 D -N)} \; \; , \label{eq:A51}
\eea
where we introduced the short-hand notations
\bea
\nu_{ijk\ldots}&=& \nu_i + \nu_j + \nu_k + \ldots \nonumber\\
N&=& \nu_{12345} \;\; .
\eea
In the above equation~(\ref{eq:A51}), $\eta>0$ is an infinitesimal quantity that indicates the way in which the analytical continuation
has to be performed in the case $q^2>0$.

In the special case in which all $\nu_i$ are equal to unity, the result simplifies
considerably. Defining a pre-factor $\ESGamma$ as
\be
\dps \ESGamma = \frac{1}{(4\pi)^{D/2} \, \Gamma(1-\eps)} \;\; ,
\ee
we have
\bea
\dps A_{5,1}[\nu_i=1]&=& i \, \ESGamma^3 \, \lek -\lp q\rp^2-i \, \eta \rek^{1-3\, \eps} \frac{\Gamma^6(1-\eps) \, \Gamma(2\,\eps) \,
\Gamma(3\,\eps) \, \Gamma(1-3\,\eps)}{(1-2\,\eps) \, (2-3\,\eps) \, \Gamma(3-4\,\eps)} \; \; .
\eea

In the next diagram, $A_{5,2}$, the power of the upper sloped propagator is labelled by $\nu_1$. $\nu_2$ and
$\nu_3$ are the powers of the propagators of the lower bubble insertion, whereas $\nu_4$ and $\nu_5$ are associated with the propagators
of the vertical bubble. From the form of the diagram we can read off that the result will be symmetric in $\{\nu_2,\nu_3\}$ as well as in
$\{\nu_4,\nu_5\}$. It reads

\bea
\dps A_{5,2}[\nu_i] &=& \loopint{k}\loopint{l}\loopint{r} \; \; \frac{1}{\lek(k+p_1)^2\rek^{\nu_1} \, \lek (l-k+p_2)^2\rek^{\nu_2}  
\, \lek l^2\rek^{\nu_3} \, \lek (k+r)^2 \rek^{\nu_4} \, \lek r^2\rek^{\nu_5}} \nnb \\
&&\nnb \\
&=&\frac{i \, (-1)^{1-N}}{(4\pi)^{3D/2}}  \, \lek -\lp q\rp^2-i \, \eta \rek^{3D/2-N} \; \frac{\Gamma(\ts{D}{2}-\nu_2) \,
\Gamma(\ts{D}{2}-\nu_3) \, \Gamma(\ts{D}{2}-\nu_4) \, \Gamma(\ts{D}{2}-\nu_5)}{\Gamma(\nu_1) \, \Gamma(\nu_2) \, \Gamma(\nu_3) \,
\Gamma(\nu_4)\, \Gamma(\nu_5)}
\nnb\\
&&\nnb \\
&& \times \, \frac{\Gamma(N-\ts{3 \, D}{2}) \, \Gamma(D-\nu_{145}) \, \Gamma(\nu_{45}-\ts{D}{2}) \,
\Gamma(\ts{3\,D}{2}- N + \nu_{1})}{\Gamma(D-\nu_{23}) \, \Gamma(D-\nu_{45}) \, \Gamma(2 D -N)} \; \; . \label{eq:A52}
\eea
Again, the case in which all $\nu_i$ are equal to unity is much simpler, namely
\bea
\dps A_{5,2}[\nu_i=1]&=& - i \, \ESGamma^3 \, \lek -\lp q\rp^2-i \, \eta \rek^{1-3\, \eps} \frac{\Gamma^7(1-\eps) \, \Gamma(\eps) \,
\Gamma(3\,\eps) \, \Gamma(1-3\,\eps)}{(1-2\,\eps) \, \Gamma(2-2\,\eps) \, \Gamma(3-4\,\eps)} \; \; .
\eea

The last diagram with two bubble insertions is $A_{6,1}$. Again, $\nu_1$ and $\nu_2$ are the powers of the sloped propagators. $\nu_3$
and $\nu_4$ form the powers of the upper bubble insertion, whereas $\nu_5$ and $\nu_6$ are given to the lower one. The diagram also
shows several symmetries, namely in $\{\nu_1,\nu_2\}$, $\{\nu_3,\nu_4\}$, $\{\nu_5,\nu_6\}$, and, in addition, in
$\{\{\nu_3,\nu_4\},\{\nu_5,\nu_6\}\}$. One finds
\bea
\dps A_{6,1}[\nu_i] &=& \loopint{k}\loopint{l}\loopint{r} \; \; \frac{1}{\lek(k+p_1)^2\rek^{\nu_1} \, \lek (k-p_2)^2\rek^{\nu_2}} \nnb \\
&&\nnb \\
&& \hs{180} \times \, \frac{1}{\lek l^2\rek^{\nu_3} \, \lek (l+k)^2 \rek^{\nu_4} \, \lek r^2\rek^{\nu_5} \, \lek (r+k)^2
\rek^{\nu_6}}\nnb \\
&&\nnb \\
&=&\frac{i \, (-1)^{1-N}}{(4\pi)^{3D/2}}  \, \lek -\lp q\rp^2-i \, \eta \rek^{3D/2-N} \; \frac{\Gamma(\ts{D}{2}-\nu_3) \,
\Gamma(\ts{D}{2}-\nu_4) \, \Gamma(\ts{D}{2}-\nu_5) \, \Gamma(\ts{D}{2}-\nu_6)}{\Gamma(\nu_1) \, \Gamma(\nu_2) \, \Gamma(\nu_3) \,
\Gamma(\nu_4)\, \Gamma(\nu_5)\, \Gamma(\nu_6)}\nnb\\
&&\nnb \\
&& \times \, \frac{\Gamma(N-\ts{3 \, D}{2}) \, \Gamma(\nu_{34}-\ts{D}{2}) \, \Gamma(\nu_{56}-\ts{D}{2}) \,
\Gamma(\ts{3\,D}{2}- N + \nu_{1}) \, \Gamma(\ts{3\,D}{2}- N + \nu_{2})}{\Gamma(D-\nu_{34}) \, \Gamma(D-\nu_{56}) \, \Gamma(2 D -N)} \;
\; , \label{eq:A61}
\eea
where this time we have
$N= \nu_{123456} \; .$

Finally, we again give the result for the case in which all $\nu_i$ are equal to unity.
\bea
\dps A_{6,1}[\nu_i=1]&=& - i \, \ESGamma^3 \, \lek -\lp q\rp^2-i \, \eta \rek^{-3\, \eps} \frac{\Gamma^7(1-\eps) \, \Gamma^2(\eps) \,
\Gamma(3\,\eps) \, \Gamma^2(1-3\,\eps)}{\Gamma^2(2-2\,\eps) \, \Gamma(2-4\,\eps)} \; \; .
\eea

Since from now on the diagrams will become more complicated, we restrain ourselves to the case in which the powers of all
propagators are equal to unity. The remaining three diagrams to be considered are $A_{6,3}$, $A_{7,1}$, and $A_{7,2}$, each of which
contains a single bubble insertion.
After integrating out the bubble insertion we are left with an effective 
two-loop diagram with one propagator less. However, one of
the propagators in the effective two-loop graph will carry a power that is different from unity. The two-loop crossed vertex graphs with powers different from
unity were discussed previously in~\cite{moch6}.

While computing the effective two-loop diagrams, it turns out that, 
after integrating over the loop momenta, all integrals over
Feynman parameters can be carried out in a closed form. 
The respective results contain $\Gamma$-functions in combination with
hypergeometric functions of unit argument. We used the aforementioned {\tt Mathematica} package {\tt
HypExp}~\cite{hypexp} for expanding the all-order results into their respective Laurent series expansions about $\eps=0$. The
explicit result for $A_{6,3}$ reads
\bea
\dps A_{6,3} &=& \loopint{k}\loopint{l}\loopint{r} \; \; \frac{1}{\lp k\rp^2 \, \lk k-q\rk^2 \, \lk k-l\rk^2  \,
\lk l-p_1\rk^2 \, \lk r-l\rk^2\, \lp r \rp^2} \nnb \\
&&\nnb \\
&=& -i \, \ESGamma \, \frac{\Gamma(\eps)\,\Gamma^3(1-\eps)}{\Gamma(2-2\,\eps)} \cdot I_5(\eps)
\eea
with
\bea
\dps I_{5}(\alpha) &=& - (-1)^{\alpha} \, \loopint{k}\loopint{l}\; \; \frac{1}{\lp k\rp^2 \, \lk k-q\rk^2 \, \lk k-l\rk^2  \,
\lk l-p_1\rk^2 \, \lek l^2\rek^{\alpha}} \nnb \\
&&\nnb \\
&=& \ESGamma^2 \, \lek -\lp q\rp^2-i \, \eta \rek^{-\alpha-2\, \eps} \frac{\Gamma^3(1-\eps)\, \Gamma(1-\alpha-\eps)\,
\Gamma(1-\alpha-2\,\eps)}{\Gamma(\alpha) \, \Gamma(2-\alpha-2\,\eps) \, \Gamma(2-\alpha-3\,\eps)} \nnb \\
&&\nnb \\
&& \times \bigg[ \frac{\Gamma(1-\alpha-2\,\eps) \, \Gamma(\alpha+2\,\eps) \, \Gamma(\alpha+\eps) \, \Gamma(\alpha) \,
\Gamma(1-\alpha-\eps)}{\Gamma(1-\eps)} \nnb \\
&&\nnb \\
&& \hs{20} + \, \frac{\Gamma(\alpha+2\,\eps-1)\,\Gamma(1-\eps)}{(1-2\,\eps)} \;
\pFq{3}{2}{1,1-\eps,1-2\,\eps}{2-2\,\eps,2-\alpha-2\,\eps}{1}\bigg] \; .\nnb\\ \label{eq:I5}
\eea
Substituting $\alpha=\eps$ in  Eq.~(\ref{eq:I5}) leads to the following series expansion for $A_{6,3}$
\bea
\dps A_{6,3} &=& i \, \ESGamma^3 \lek -\lp q\rp^2-i \, \eta \rek^{-3\, \eps}\nnb\\
&&\nnb \\
&& \times \, \lek  - \frac{1}{6\,{\eps}^3} - \frac{3}{2\,{\eps }^2} - \lk
   \frac{55}{6} + \frac{{\pi }^2}{6}\rk\frac{1}{\eps } -\frac{95}{2}  - \frac{3\,{\pi }^2}{2}+ 
  \frac{17\,\zeta_3}{3} \rp\nnb \\
&&\nnb \\
&& \hs{18} + \lk - \frac{1351}{6}  - 
     \frac{55\,{\pi }^2}{6} - \frac{{\pi }^4}{90} + 
     51\,\zeta_3 \rk \, \eps \nnb \\
&&\nnb \\
&& \hs{18}  + \lk - \frac{2023}{2}  - 
     \frac{95\,{\pi }^2}{2} - \frac{{\pi }^4}{10} + 
     \frac{935\,\zeta_3}{3} + 
     \frac{10\,{\pi }^2\,\zeta_3}{3} + 
     65\,\zeta_5 \rk \, \eps^2\, \nnb \\
&&\nnb \\
&& \hs{18} + \lk - \frac{26335}{6}   - 
     \frac{1351\,{\pi }^2}{6} - \frac{11\,{\pi }^4}{18} + 
     \frac{7\,{\pi }^6}{54} \rp \nnb \\
&&\nnb \\
&& \hs{35} \lp\lp  + 1615\,\zeta_3 + 
     30\,{\pi }^2\,\zeta_3 - 
     \frac{268\,\zeta_3^2}{3} + 585\,\zeta_5
     \rk \, \eps^3 + {\cal O} (\eps^4)\rek \; \; .
\eea
The above Eq.~(\ref{eq:I5}) can be used for two other cross-checks. 
First, we can consider the limit $\alpha \to 0$. This is
done by setting $\alpha=\xi \, \eps$, followed by the series expansion in $\eps$. Finally, we set $\xi=0$. The result has to coincide --
up to a global sign -- with the series expansion of the 
two-loop integral $A_4$ of Eq.~(4) in Ref.~\cite{ghm}. The second check is
performed by the limit $\alpha \to 1$, in which case we have 
to find the result for the two-loop five propagator integral that is obtained
from $A_{6,3}$ by removing the bubble. 
Both checks were found to be fulfilled on the level of the series expansions.
The calculation of $I_5(\eps)$ by sector decomposition 
provided an additional check.

We now proceed with the integral $A_{7,1}$, which assumes the form
\bea
\dps A_{7,1} &=& \loopint{k}\loopint{l}\loopint{r} \; \; \frac{1}{\lp r\rp^2 \, \lk r-k\rk^2 \, \lk k-q\rk^2  \,
\lk k-l\rk^2 \, \lk k-l-p_2\rk^2\, \lp l \rp^2 \, \lk l-p_1\rk^2} \nnb \\
&&\nnb \\
&=& -i \, \ESGamma \, \frac{\Gamma(\eps)\,\Gamma^3(1-\eps)}{\Gamma(2-2\,\eps)} \cdot I_6(\eps)
\eea
with
\bea
\dps I_{6}(\alpha) &=& - (-1)^{\alpha} \, \loopint{k}\loopint{l}\; \; \frac{1}{\lek k^2\rek^{\alpha}\, \lk k-q\rk^2 \, \lk k-l\rk^2  \,
\lk k-l-p_2\rk^2 \, \lp l \rp^2 \, \lk l-p_1\rk^2} \nnb \\
&&\nnb \\
&=& \ESGamma^2 \, \lek -\lp q\rp^2-i \, \eta \rek^{-1-\alpha-2\, \eps} \Gamma^2(1-\eps)\, \Gamma^2(-\eps) \nnb \\
&&\nnb \\
&& \times \bigg[ \frac{\Gamma(-\eps) \, \Gamma(2\,\eps)}{2 \, \Gamma(1-3\,\eps)} \;
\pFq{4}{3}{\alpha,1-\alpha-4\,\eps,1-\eps,-2\,\eps}{1-3\,\eps,1-2\,\eps,1-2\,\eps}{1}\nnb \\
&& \hs{20} + \, \frac{\Gamma(1-2\,\eps)\,\Gamma(1-\alpha-2\,\eps)\,\Gamma(2+\eps)\,\Gamma(\alpha+2\,\eps)}{\Gamma(\alpha)\,
\Gamma(2-\eps)\,\Gamma(1-\alpha-4\,\eps)} \nnb\\
&& \hs{35} \times \, \pFq{4}{3}{1,1,1-2\,\eps,2+\eps}{2,2,2-\eps}{1} \nnb \\
&& \hs{20} - \, \frac{2 \, \Gamma(-2\,\eps)\,\Gamma(1+\alpha+2\,\eps)\,\Gamma(2+\eps)\,\Gamma(1-\alpha-2\,\eps)}{\Gamma(\alpha)\,
\Gamma(2-\eps)\,\Gamma(1-\alpha-4\,\eps)} \nnb\\
&& \hs{35} \times \, \pFq{4}{3}{1,1,1+\alpha+2\,\eps,2+\eps}{2,2,2-\eps}{1} \nnb \\
&& \hs{20} - \, \frac{\Gamma(\alpha+2\,\eps)\,\Gamma(2-\alpha-\eps)}{\lk1-\alpha-2\,\eps\rk^2\,\Gamma(\alpha)\,
\Gamma(2-\alpha-3\,\eps)} \nnb\\
&& \hs{-15} \times \,
\pFq{4}{3}{1,1-\alpha-2\,\eps,1-\alpha-4\,\eps,2-\alpha-\eps}{2-\alpha-2\,\eps,2-\alpha-2\,\eps,2-\alpha-3\,\eps}{1} \nnb \\
&& \hs{20} + \, \frac{\Gamma(-2\,\eps)\, \Gamma(1+2\,\eps)\,\Gamma(2+\eps)\,\Gamma(1+\alpha+2\,\eps)\,
\Gamma(2-\alpha-2\,\eps)}{\Gamma(\alpha)\,\Gamma(1-\alpha-4\,\eps)\,\Gamma(2-\eps)\,\Gamma(2+2\,\eps)} \nnb\\
&& \hs{35} \times \,
\pFq{5}{4}{1,1,2-\alpha-2\,\eps,1+\alpha+2\,\eps,2+\eps}{2,2,2-\eps,2+2\,\eps}{1} \bigg]\; . \label{eq:I6}
\eea
Again, we have to set $\alpha=\eps$ in  Eq.~(\ref{eq:I6}) in order to obtain the series expansion for $A_{7,1}$. It reads
\bea
\dps A_{7,1} &=& i \, \ESGamma^3 \lek -\lp q\rp^2-i \, \eta \rek^{-1-3\, \eps}\nnb\\
&&\nnb \\
&& \times \, \lek  \frac{1}{4\,\eps^5} + \frac{1}{2\,\eps^4} + \lk 1 - \frac{\pi^2}{6}\rk\frac{1}{\eps^3}
+ \lk 2 - \frac{\pi^2}{3} - 10\,\zeta_3\rk\frac{1}{\eps^2} \rp\nnb \\
&&\nnb \\
&& \hs{18} + \lk 4 - \frac{2\,{\pi }^2}{3} - \frac{11\,{\pi }^4}{45} - 20\,\zeta_3\rk \, \frac{1}{\eps} \nnb \\
&&\nnb \\
&& \hs{18} + \lk 8 - \frac{4\,{\pi }^2}{3}
- \frac{22\,{\pi }^4}{45} - 40\,\zeta_3 + \frac{14\,{\pi }^2\,\zeta_3}{3} - 88\,\zeta_5 \rk  \nnb \\
&&\nnb \\
&& \hs{18} + \lk 16 - \frac{8\,{\pi }^2}{3} - \frac{44\,{\pi }^4}{45} - 
    \frac{943\,{\pi }^6}{7560} - 80\,\zeta_3 \rp \nnb \\
&&\nnb \\
&& \hs{35} \lp\lp + \frac{28\,{\pi }^2\,\zeta_3}{3} + 196\,\zeta_3^2 - 176\,\zeta_5  \rk \eps + {\cal O} (\eps^2)\rek \; \; .
\eea
The integral $I_6(\alpha)$ provides another cross check since for $\alpha=1$ we have to reproduce the integral $A_6$ of
Eq.~(5) in Ref.~\cite{ghm}. This we checked to be the case on the level of the series expansion.

As we proceed, the expressions for the integrals become more and more lengthy. The result for the integral $A_{7,2}$ reads
\bea
\dps A_{7,2} &=& \loopint{k}\loopint{l}\loopint{r} \; \; \frac{1}{\lp k\rp^2 \, \lk k-q\rk^2  \,
\lk l-p_1\rk^2 \,\lk k-l\rk^2 \, \lk k-l-p_2\rk^2\, \lp r \rp^2 \, \lk r-l\rk^2} \nnb \\
&&\nnb \\
&=& -i \, \ESGamma \, \frac{\Gamma(\eps)\,\Gamma^3(1-\eps)}{\Gamma(2-2\,\eps)} \cdot J_6(\eps)
\eea
with
\bea
\dps J_{6}(\alpha) &=& - (-1)^{\alpha} \, \loopint{k}\loopint{l}\; \; \frac{1}{\lp k\rp^2 \, \lek l^2\rek^{\alpha} \, \lk k-q\rk^2  \,
\lk l-p_1\rk^2 \,\lk k-l\rk^2 \, \lk k-l-p_2\rk^2} \nnb \\
&&\nnb \\
&=& \ESGamma^2 \, \lek -\lp q\rp^2-i \, \eta \rek^{-1-\alpha-2\, \eps} \Gamma(1-\eps)\, \Gamma(-\eps) \, \Gamma(1-\alpha-\eps)\nnb \\
&&\nnb \\
&& \times \bigg[ -\frac{\Gamma(1-\alpha-2\,\eps) \,\Gamma(\alpha+\eps)
\,\Gamma(\alpha+2\,\eps)\,\Gamma(1-\eps)\, \Gamma(-\eps)\, \Gamma^2(\eps)}{4\, \Gamma(\alpha) \, \Gamma(1-\alpha-4\,\eps) \,
\Gamma(2\,\eps)} \nnb \\
&&\nnb \\
&& \hs{20} +\frac{\Gamma(1-\alpha-2\,\eps) \,\Gamma(\alpha+\eps)
\,\Gamma(\alpha+2\,\eps)\, \Gamma(-2\,\eps)}{\eps\, \Gamma(\alpha) \, \Gamma(1-\alpha-4\,\eps)} \nnb \\
&&\nnb \\
&& \hs{20} -\frac{\Gamma^2(1-\alpha-2\,\eps) \,\Gamma(\alpha+\eps)
\,\Gamma^2(\alpha+2\,\eps)\, \Gamma(\alpha+4\,\eps)}{\Gamma(\alpha) \, \Gamma(1+2\,\eps)} \nnb \\
&&\nnb \\
&& \hs{20} -\frac{\Gamma(1-\alpha-2\,\eps)\, \Gamma(1-\eps) \,\Gamma(\alpha+\eps)
\,\Gamma(\alpha+2\,\eps-1)}{\Gamma(\alpha) \, \Gamma(2-\alpha-3\,\eps)} \nnb \\
&& \hs{35} \times \, \pFq{3}{2}{1,1-\alpha-2\,\eps,1-\alpha-4\,\eps}{2-\alpha-2\,\eps,2-\alpha-3\,\eps}{1}\nnb \\
&& \hs{20} +\frac{\Gamma(\alpha-1)\, \Gamma(1-\alpha-2\,\eps)\, \Gamma(1-\eps) \,\Gamma(-2\,\eps)
\,\Gamma(1+\eps)\,\Gamma(1+2\,\eps)}{\Gamma(\alpha) \, \Gamma(1-\alpha-4\,\eps)\,\Gamma(2-\alpha-\eps)} \nnb \\
&& \hs{35} \times \, \pFq{3}{2}{1-\alpha,1+\eps,1+2\,\eps}{2-\alpha,2-\alpha-\eps}{1}\nnb \\
&& \hs{20} +\frac{\Gamma(1-\eps) \,\Gamma(-2\,\eps)
\,\Gamma(1+\eps)\,\Gamma(1+2\,\eps)\,\Gamma(\alpha+\eps)\,\Gamma(\alpha+2\,\eps)}{(\alpha+4\,\eps)\,\Gamma(\alpha) \,
\Gamma(1-\alpha-3\,\eps)\,\Gamma(1+\alpha+3\,\eps)} \nnb \\
&& \hs{35} \times \, \pFq{3}{2}{1+\eps,1+2\,\eps,\alpha+4\,\eps}{1+\alpha+3\,\eps,1+\alpha+4\,\eps}{1}\nnb \\
&& \hs{20} +\frac{\Gamma(1-2\,\eps) \, \Gamma(1-\alpha-2\,\eps) \, \Gamma(1-\eps)
\,\Gamma(1+\alpha+\eps)\,\Gamma(\alpha+2\,\eps)}{\,\Gamma(1+\alpha) \,
\Gamma(1-\alpha-4\,\eps)\,\Gamma(2-\eps)} \nnb \\
&& \hs{35} \times \, \pFq{4}{3}{1,1,1-2\,\eps,1+\alpha+\eps}{2,1+\alpha,2-\eps}{1}\nnb \\
&& \hs{20} +\frac{\Gamma^2(1-\eps)\,\Gamma(2\,\eps)\,\Gamma(\alpha+\eps)\,\Gamma(\alpha+2\,\eps)}{(\alpha+2\,\eps)\,\Gamma(\alpha) \,
\Gamma(1-\alpha-3\,\eps)\,\Gamma(1+\alpha+\eps)\,\Gamma(1+2\,\eps)} \nnb \\
&& \hs{35} \times \, \pFq{4}{3}{1,1,1-\eps,\alpha+2\,\eps}{1-2\,\eps,1+\alpha+\eps,1+\alpha+2\,\eps}{1}\nnb \\
&& \hs{20} +\frac{\Gamma(1-\alpha-2\,\eps)\,\Gamma^2(1-\eps)\,\Gamma(\alpha+2\,\eps-1)}
{\Gamma(2-\alpha-3\,\eps)\,\Gamma(1-2\,\eps)}\nnb \\
&& \hs{20} \times \, \pFq{4}{3}{1,1-\alpha-4\,\eps,1-\alpha-2\,\eps,1-\eps}{2-\alpha-2\,\eps,2-\alpha-3\,\eps,1-2\,\eps}{1}\nnb \\
&& \hs{20} +\frac{\alpha \, \Gamma(1-\alpha-2\,\eps)\,\Gamma(-2\,\eps)\,\Gamma(\alpha+\eps)\,\Gamma(\alpha+2\,\eps)}
{\Gamma(\alpha)\,\Gamma(1-\alpha-4\,\eps)}\; \pFq{3}{2}{1,1,1+\alpha}{2,2}{1}\nnb \\
&& \hs{20} -\frac{\Gamma(1-\alpha-2\,\eps)\,\Gamma(-2\,\eps)\,\Gamma(\alpha+\eps)\,\Gamma(1+\alpha+2\,\eps)}
{\Gamma(\alpha)\,\Gamma(1-\alpha-4\,\eps)}\nnb \\
&& \hs{35} \times \, \pFq{3}{2}{1,1,1+\alpha+2\,\eps}{2,2}{1} \, \bigg]\; . \label{eq:J6}
\eea
Details about the calculation of Eq.~(\ref{eq:J6}) can be found in Ref.~\cite{bachelorcedric}. Useful formulas that got applied at
intermediate steps were taken from Refs.~\cite{thebook1,thebook2,wolframweb}. Setting $\alpha=\eps$ leads to the
following series expansion of $A_{7,2}$,
\bea
\dps A_{7,2} &=& i \, \ESGamma^3 \lek -\lp q\rp^2-i \, \eta \rek^{-1-3\, \eps}\nnb\\
&&\nnb \\
&& \times \, \lek  \frac{\pi^2}{12\,\eps^3} + \lk \frac{\pi^2}{6} + 2 \,\zeta_3 \rk\frac{1}{\eps^2}
+ \lk \frac{\pi^2}{3}+\frac{83 \,\pi^4}{720}+4\,\zeta_3\rk\frac{1}{\eps} \rp\nnb \\
&&\nnb \\
&& \hs{18} + \lk  \frac{2\,\pi^2}{3}
+ \frac{83\,\pi^4}{360} + 8\,\zeta_3 - \frac{5\,\pi^2\,\zeta_3}{3} + 15\,\zeta_5 \rk  \nnb \\
&&\nnb \\
&& \hs{18} + \lk \frac{4\,\pi^2}{3} + \frac{83\,\pi^4}{180} + 
    \frac{2741\,\pi^6}{90720} + 16\,\zeta_3 \rp \nnb \\
&&\nnb \\
&& \hs{35} \lp\lp - \frac{10\,\pi^2\,\zeta_3}{3} - 73\,\zeta_3^2 + 30\,\zeta_5  \rk \eps + {\cal O} (\eps^2)\rek \; \; .
\eea
We finally state that the expression~(\ref{eq:J6}) for $J_{6}(\alpha)$ can again be used for several cross  checks. First, in the limit
$\alpha \to 1$ we have to obtain the same result as for $A_6$ of Eq.~(5) in Ref.~\cite{ghm} or $I_6(1)$ of Eq.~(\ref{eq:I6}). The check is
done by first considering $\alpha = 1 + \chi \, \eps$ in (\ref{eq:J6}) followed by a subsequent expansion in $\eps$. In the end, the limit
$\chi \to 0$ is carried out. A second check is provided by the limit $\alpha \to 0$. We again set $\alpha = \eta \, \eps$ and carry out
the series expansion, followed by letting $\eta \to 0$. The result has to be the same -- up to a global sign -- as the series expansion
of $I_5(1)$ of Eq.~(\ref{eq:I5}). All checks have been verified on the level of the respective Laurent series.

As mentioned earlier, the coefficients of the Laurent series are real numbers.
Therefore the method of sector decomposition
is particularly well suited to compute the coefficients numerically, 
thereby providing the most important check of our analytical
findings. 

\section{Conclusions} 
\label{sec:conc}
In this letter, we identified and classified the master integrals 
required for a calculation of the massless three-loop quark and 
gluon form factors. In addition to three-loop two-point functions and 
products of one-loop and two-loop integrals, we identified 16 genuine 
three-loop vertex integrals, which are displayed in 
Figure~\ref{fig:3loopdiagrams}. Among these, six integrals are so-called 
insertion graphs, containing  a
bubble insertion into a one-loop or two-loop vertex graph. We computed the
master integrals for these insertion graphs analytically in a 
closed form which is exact to all orders in $\e$, containing 
$\Gamma$-functions and hypergeometric functions. Laurent series 
expansions were subsequently obtained using the {\tt HypExp}-package. 
All Laurent series expansions were verified independently using 
sector decomposition to determine the expansion coefficients numerically. 

The remaining ten master integrals do not contain subtopologies which 
would allow us to relate them to two-loop integrals. Their analytical 
computation may not be possible using Feynman parameters, but
appears feasible with modern loop-integral techniques\,\cite{smirnovbook}, 
such as Mellin-Barnes integration~\cite{smirnov}. 
Using sector decomposition, 
their Laurent expansion 
can be obtained in a straightforward manner.

\section*{Acknowledgements}
This research was supported by the Swiss National Science Foundation
(SNF) under contract 200020-109162.

\end{document}